\documentclass[prl,a4paper]{revtex4}
\usepackage{comment}
\usepackage{color}
\usepackage{amsmath}
\usepackage{amssymb}
\usepackage{graphicx}
\usepackage{braket} 
\usepackage{esint}
\usepackage{adjustbox} 
\usepackage[utf8]{inputenc}

\usepackage[unicode=true,pdfusetitle,
bookmarks=true,bookmarksnumbered=false,bookmarksopen=false,
breaklinks=true,pdfborder={0 0 1},backref=false,colorlinks=true]
{hyperref}
\hypersetup{
	linkcolor=red,urlcolor=blue,citecolor=blue,anchorcolor=blue} 
\usepackage{ulem} 

\usepackage{dcolumn}\usepackage{bm}

\makeatother
 
\begin{document}
%\preprint{}
\title{Retroreflection and diffraction of a Bose-Einstein condensate by evanescent standing wave potential}
\author{Javed Akram \thanks{Corresponding author}  }\email{javedakram@daad-alumni.de  }
\affiliation{Department of Physics, COMSATS University Islamabad, 45550 Pakistan}
\author{Khan Qasim} % \email{qasimkhan@seu.edu.cn}
\author{Lei Wei} %\email{lw@seu.edu.cn}
\affiliation{Joint International Research Laboratory of Information Display and Visualization, School of Electronic Science and Engineering, Southeast University, Nanjing, Jiangsu 210096, China} 
\date{\today}

\begin{abstract}
The characteristic of the angular distributions of accelerated Bose-Einstein condensate (BEC) atoms incidence on the surface is designed using the mathematical modeling method. Here, we proposed the idea to study retroreflection and diffraction of a BEC from an evanescent standing wave potential (ESWP). The ESWP is formed by multiple reflections of the laser beam from the surface of the prism under the influence of gravity. 
After BEC's reflection and diffraction, the so-called BEC's density rainbow patterns develop due to the interference which depends on the surface structure which we model with the periodic decaying evanescent field.  The interaction of accelerated bosonic atoms with a surface can help to demonstrate  surface structures or to determine surface roughness, or to build future high spatial resolution and high sensitivity magnetic-field sensors in two-dimensional systems.

\end{abstract}


\maketitle

% Force line breaks with \\

% \altaffiliation[Also at ]{ Department, XYZ University.}%Lines break automatically or can be forced with \\

%\author{Javed Akram}
%\affiliation{



\section{Introduction} \label{Sec1}
The thermal and quantum fluctuations started to play an important role in ultra-cold atoms at low dimensions, therefore low dimensions got much attention in the past few  
years \cite{PhysRevLett.95.190403,PhysRevA.80.021602,Nanofiber,PhysRevA.93.023606,PhysRevA.93.033610,Akram_2018}.  
Under such circumstances also the influence of gravity has to be taken into account. Atomic mirrors has been constructed to study the influence of reflection and diffraction of atomic beams and cold atomic clouds, these atomic mirrors were made by using repulsive evanescent waves \cite{COOK1982258,Liston1995}. 
 Cohen-Tannoudji studied a new atomic cavity consisting of a single horizontal concave mirror placed in a gravity, where the gravity acts as a second mirror closing the cavity \cite{Wallis1992,Wallis_1996}.  
In these kind of atomic mirrors, inherent losses of atoms are present due to quantum and thermal fluctuations, whoever the loses can be reduced by using the higher detuning of atomic resonance frequency and evanescent wave 
 \cite{PhysRevLett.71.3083,PhysRevLett.74.4972}. 

The dynamics of a quantum particle bouncing on a hard 
surface under the influence of gravity referred as a quantum bouncing ball (QBB) \cite{Gea-Banacloche-1999}, is a notable example of a quantum 
mechanical system showing collapses and revivals of the QBB, which is a clear signature of quantum interference without classical 
correspondence \cite{ROBINETT20041}. Lately, this kind of system open the opportunities to study the dynamics of atoms in the modulated evanescent wave (EW), 
such as localization \cite{PhysRevA.58.4779,Akram-2009} or a coherent acceleration \cite{SAIF2005207,Akram-Khalid-2009} of atoms, strongly depending on the system parameters and 
on the chosen initial condensations, which is signature of chaos in this kind of system. 
During the interaction of atoms with surface, a good loading scheme and trap geometry become important for studying the dynamics of the BEC in gravito-optical surface trap (GOST).
The experimental group of Rudi Grimm demonstrated both the loading of Cesium atoms \cite{PhysRevLett.79.2225,GRIMM200095} 
and the subsequent creation of a BEC in a two-dimensional GOST \cite{Domokos2001,PhysRevLett.92.173003}. 
As the QBB problem was for long time regarded as a main pedagogical interest 
\cite{PhysRevA.58.4779,Akram-2009,SAIF2005207,Akram-Khalid-2009}, 
recently many experimental group demonstrated the QBB using cold atoms \cite{PhysRevLett.71.3083,DOWLING19961,PhysRevLett.82.468,PhysRevLett.83.3577} and by owning the formation of a 
QBB the energy levels of the bouncing neutrons are quantized \cite{Nesvizhevsky-2002}. 
Colombe et al. studied the scheme for loading a $^{87}$Rb BEC into a two-dimensional evanescent light trap 
and observe the diffraction of a BEC in the time domain \cite{Colombe_2003,PhysRevA.72.061601}. Afterwards Perrin et at. studied the diffuse reflection of a BEC from a rough evanescent wave mirror \cite{Perrin_2006}. 
\begin{figure}[ht]
\begin{center}
\includegraphics[width=7.00in,height=2.30in]{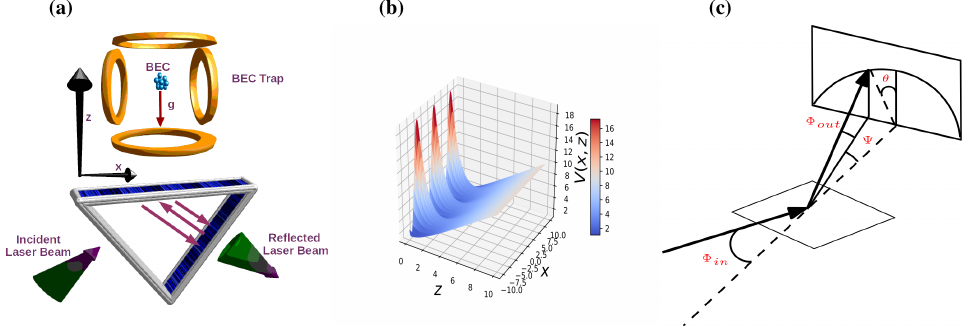}
\end{center} 
\caption{(a) A BEC is trapped in a trap at a certain height above a dielectric slab.
A ESWP created by the total internal reflection of the incident
laser beam from the surface of the dielectric, which serves as a mirror for the BEC. 
(b) A two-dimensional plot of the dimensionless potential, where the potential is periodic along the x-axis and along the z-axis the potential of the system is linear due to the dimensionless gravity. 
Gravity and the standing evanescent wave field form a cavity for the atomic de
Broglie waves. The BEC atoms undergo a bounded motion in this gravitational cavity. (c) Sketch for the scattering of   the BEC atoms from a surface under the incident angle $\phi_{in}$ and the semi-circle indicate the elastic scattering.  }
\label{fig1}
\end{figure}
Previously, we have studied the behavior of Cs BEC in a quasi-1D GOST, we developed an analytical approximate solution of GOST and compared with numerical simulations and with Innsbruck experimental data \cite{Akram_2016}. The analytical solution of this complicated problem is only possible around GOST minimum harmonic potential  \cite{Akram_2016}. However, the system gets quite complected when the anharmonicities started to play the role, therefore, it is not possible to solve this research problem analytically. Actually these anharmonicities are an integral part of the system due to its geometry \cite{Perrin_2006}. In this scientific report, we report reflecting of BEC from counter-propagating evanescent waves, i.e., an evanescent standing wave potential (ESWP). The required optical 
wave-field can be produced by totally internally reflecting a laser beam at the surface of a refractive medium and then retro-reflecting the 
light back along its original path as shown in Fig. \ref{fig1}. 
The evanescent field decreases exponentially in the direction normal to the surface and is modulated sinusoidal along the 
surface as depicted in Fig. \ref{fig1} \cite{DOWLING19961}.

The underlying two-dimensional model for a BEC in the ESWP trap geometry is explained in Sec. \ref{Sec2}, 
where we also provide estimates for experimentally realistic parameters. A detailed study of the reflection and diffraction of the BEC wavepacket is provided in Sec.  \ref{Sec3}.  And, in Sec.  \ref{Sec4} we give a brief summary and draw our conclusions. Finally, in Sec.  \ref{Sec5} we describe a  method and numerical technique in detail. 

\begin{figure*}
\begin{center}
\includegraphics[width=6.50in,height=3.00in]{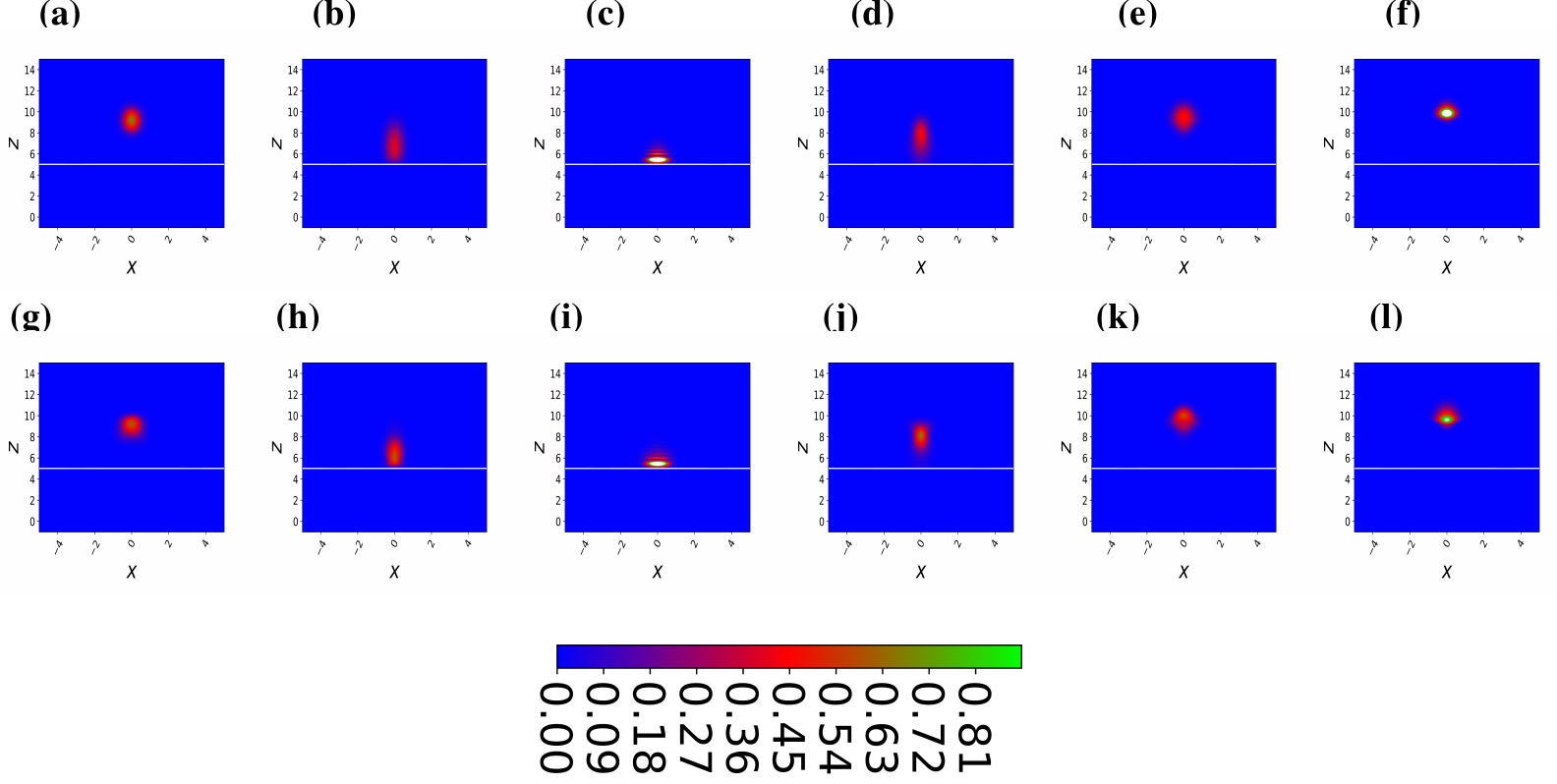}
\end{center}
\caption{(Color online) Temporal two-dimensional density plot of the BEC bouncing on the prism under the influence of gravity releases from height $z=10$, all images of the BEC cloud are captured after every $\triangle t=5$, we start with  $t=5 (a)$ and end with $t=60 (l)$. Here, the number of atoms are $N=3$,  the magnitude of the periodic potential is $\eta=0$, and the dimensionless frequency of the periodic potential defines as $ \nu= 1$.    }
\label{fig2} 
\end{figure*}

\begin{figure*}
\begin{center}
\includegraphics[width=6.50in,height=3.00in]{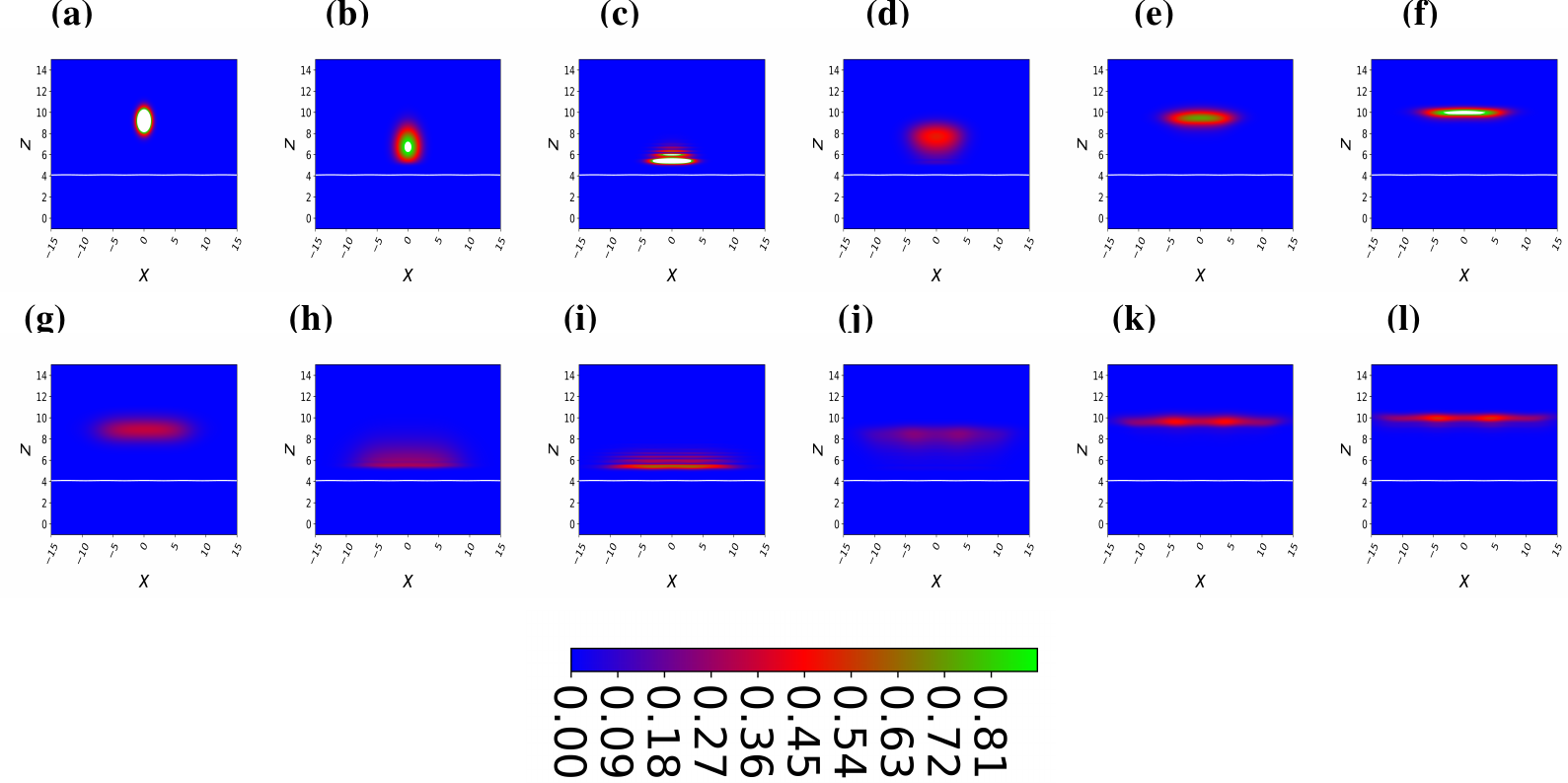}
\end{center}
\caption{(Color online)  Temporal two-dimensional density plot of the BEC bouncing on the prism under the influence of gravity releases from height $z=10$, all images of the BEC cloud are captured after every $\triangle t=5$, we start with  $t=5 (a)$ and end with $t=60 (l)$. Here, the number of atoms are $N=3$,  the magnitude of the periodic potential is $\eta=0.01$, and the dimensionless frequency of the periodic potential defines as $ \nu= 1$. }
\label{fig3}
\end{figure*}

\section{Theoretical Model} \label{Sec2}
We start working out a two-dimensional model for the dynamics of the BEC in the ESWP. Assuming that both thermal and quantum 
fluctuations can be neglected, we arrive at a mean-field theory with the two-dimensional 
time-dependent Gross-Pitaevskii  equation (GPE) \cite{PhysRevA.65.043617}
\begin{equation}
i \hbar \frac{\partial }{\partial t}\Psi (x,z,t)=\left\{ -\frac{\hbar ^{2}%
}{2m} \left( \frac{\partial ^{2}}{\partial x^{2}} +\frac{\partial ^{2}}{\partial z^{2}}  \right) +V(x,z)+G \lvert\Psi (x,z,t)\rvert
^{2}\right\} \Psi (x,z,t).  \label{Eq1}
\end{equation}
On the right-hand side the first term represents the kinetic energy of the atoms with mass $m$, the last term stands for their two-particle 
interaction, where its strength $G=  G_{3D}/(a_y \sqrt{2\pi})$, here $G_{3D}=N 4 \pi a \hbar^2/m $ is related to the s-wave scattering length $a$,  $m$ denotes a mass of the bosons, and $a_y=(\hbar /m \omega_y )^{1/2}$ defines the width of the oscillator along the $y$-axis. For this pancake configuration of the trap potential, we have in mind $\omega_y >> \omega_{\perp}$, here $\omega_y$ defines the frequency of the trap along the $y-$axis and the  $\omega_{\perp}$ defines the frequency of the radial component, 
which has to be small enough in order to ensure a quasi two-dimensional setup \cite{PhysRevA.65.043617}. In comparison to a conventional transmission grating, the atomic reflection grating has to achieve two functions: reflection and diffraction. For this purpose, we proposed here an evanescent standing wave potential. A reflection optical potential with a decaying field along the $z$-direction and a periodic modulation along the $x$-axis in the presence of gravity. This kind of periodicity can be realized by overlapping two evanescent waves with opposite wave-vectors as shown in Fig. \ref{fig1}. The BEC atoms falling off on this optical potential will experience a reflection of the $z$-component of their momentum, this reflection is only possible when their kinetic energy does not increase the potential height, this condition helps BEC to avoid to penetrate into the potential. Furthermore, 
$V(x,z)$ in Eq. (\ref{Eq1}) represents the underlying anharmonic potential energy of our system:
\begin{equation}
 V(x,z)=mgz+V_{0}\left[1+\eta \cos\left(\nu x\right)\right]e^{-\kappa z}. \label{Eq2}
\end{equation}
Here,  $g$ represents the gravitational acceleration, and $V_{0}=\Gamma \lambda^{3} I_{0} / (8\pi^{2} c \delta_{3})$ describes the strength of the 
evanescent field, where $\Gamma$ is the natural line-width of the Cs atoms, $\lambda$ defines the wavelength of the EW, 
$I_{0}$ denotes the peak intensity of the EW, and $\delta_{3}$ corresponds to the detuning frequency of the hyperfine sub-level 
$F=3$ of the Cs atom band. While, the dimensionless quantity $\eta$ is related to the magnitude of the wave vector of the standing wave \cite{Henkel-1999}, and 
$1/ \kappa=\lambda / 4\pi \sqrt{n^{2}\sin^{2}\beta -1}$ 
defines the decay length, where $n$ represents the refractive index of the medium and $\beta$ denotes the angle of incidence of the laser beam. When $\eta\rightarrow0$, the leftover anharmonic potential (\ref{Eq2}) can be approximated around its minimum 
$z^{\rm{min}}_{0}=\frac{1}{\kappa}\ln(\frac{V_{0}\kappa}{mg})$ 
with the axial frequency $\omega_{z}=\sqrt{g \kappa}$. % are depicted schematically in Fig. \ref{fig4}
In order to have a concrete set-up in mind, we will use for our analysis the following parameter values which stem from the 
GOST experiment \cite{GRIMM200095,Akram_2016}. We denote the number $N$ of Cs atoms, where the s-wave scattering length amounts to 
$a=440~\rm{a}_{0}$ with the Bohr radius $a_{0}$. The inverse decay length amounts to $\kappa=6.67\times10^{6}~\rm{m^{-1}}$ as the 
EW is produced by a far-detuned laser with wavelength $\lambda=852~\rm{nm}$
in a nearly round spot on the surface of the prism with reflective index $n=1.45$, and the angle of incidence of the laser 
light is $\beta=49.2^{o}$. The natural line width of the Cs atom is $\Gamma=2\pi \times 5.3~\rm{MHz}$,  
the detuning between the hyperfine sub-levels is $\delta_{3}=2\pi \times 1~\rm{GHz}$, and  
the peak intensity of the EW amounts to $I_{0}=9.6 \times 10^{7} ~\rm{W/m^{2}}$, so the strength of the EW is of the order 
$V_{0}=0.96\times\rm{ k_{B}}~\rm{K}$, where $\rm{k_{B}}$ is Boltzmann's constant and 
the axial harmonic frequency amounts to $\omega_{\perp}=2\pi \times 1.2 ~ \rm{kHz}$. 

In the following considerations we assume
$\omega_{y}=2\pi\times 12.0 ~ \rm{kHz}$, which fulfills the  
quasi two-dimensional condition $\omega_y >> \omega_\perp$ for the ESWP experiment. 
As the BEC does not penetrate very far into the repulsive ESWP, it is hardly 
influenced by the Van-der-Waals potential, which follows from the above values of the GOST experiment \cite{Akram_2016}. 
In order to do the numerical simulation, we use the dimensionless equation as  

\begin{equation} 
 i \frac{\partial }{\partial \tilde{t}} \tilde{\Psi} (\tilde{x}, \tilde{z},\tilde{t}) = \\
 \left\{ -\frac{k}{2} \left( 
    \frac{\partial ^{2}}{\partial \tilde{x}^{2}} +
    \frac{\partial ^{2}}{\partial \tilde{z}^{2}} \right ) +\tilde{z} + \tilde{V_0} \left[1+\eta\cos\left( \tilde{\nu} \tilde{x} \right)\right]e^{-\tilde{z}}   
    +\tilde{G} \lvert\tilde{\Psi} \left( \tilde{x}, \tilde{z},\tilde{t} \right ) \rvert
^{2}\right\} \tilde{\Psi} \left( \tilde{x}, \tilde{z},\tilde{t} \right ) , \label{Eq3}
\end{equation}

here, we consider dimensional spatial coordinates $\tilde{z}= \kappa z$ and $\tilde{x}= \kappa x$, the dimensionless time $\tilde{t}= t (m g)/\hbar \kappa$ and the two-particle dimensionless interaction strength $\tilde{G}= N 2 \sqrt{2\pi} \tilde{a} k / \tilde{a}_y$, here, $\tilde{a}=a\kappa $ being a dimensionless s-wave scattering length, $\tilde{a_y}= a_y \kappa$ is the dimensionless oscillator length along the y-axis, and $k= \hbar^2 \kappa^3 /(g m^2)$ defines the dimensionless kinetic energy constant. Additionally, we measure energies in gravitational energy $m g /\kappa$ and dimensionless frequency describe as $\tilde{\omega_z } = \omega_z \hbar \kappa  /(m g)$ and dimensionless strength of the evanescent field represents as $\tilde{V_{0}}= \kappa V_{0} / (mg)$.
By keeping in mind the last experimental values, we have the following dimensionless quantities: the dimensionless s-wave scattering length amounts to $\tilde{a}=0.033$, the dimensionless optical decaying strength is  given by 
$\tilde{V}_{0}=906$, 
the dimensionless kinetic energy strength is $\tilde{k}=0.066$,  the dimensionless radial frequency amounts to be $\tilde{\omega}_{r}=1.303$, and  
dimensionless two-particle interaction yields $\tilde{G}=0.086 ~ N$. We have normalized the wavefunction as $\int \lvert \Psi \left(x,z \right ) \rvert ^{2} dx dz =1$. To avoid complexity, we drop all parameter tildes for rest of the manuscript.

\begin{figure*}
\begin{center}
\includegraphics[width=6.50in,height=3.00in]{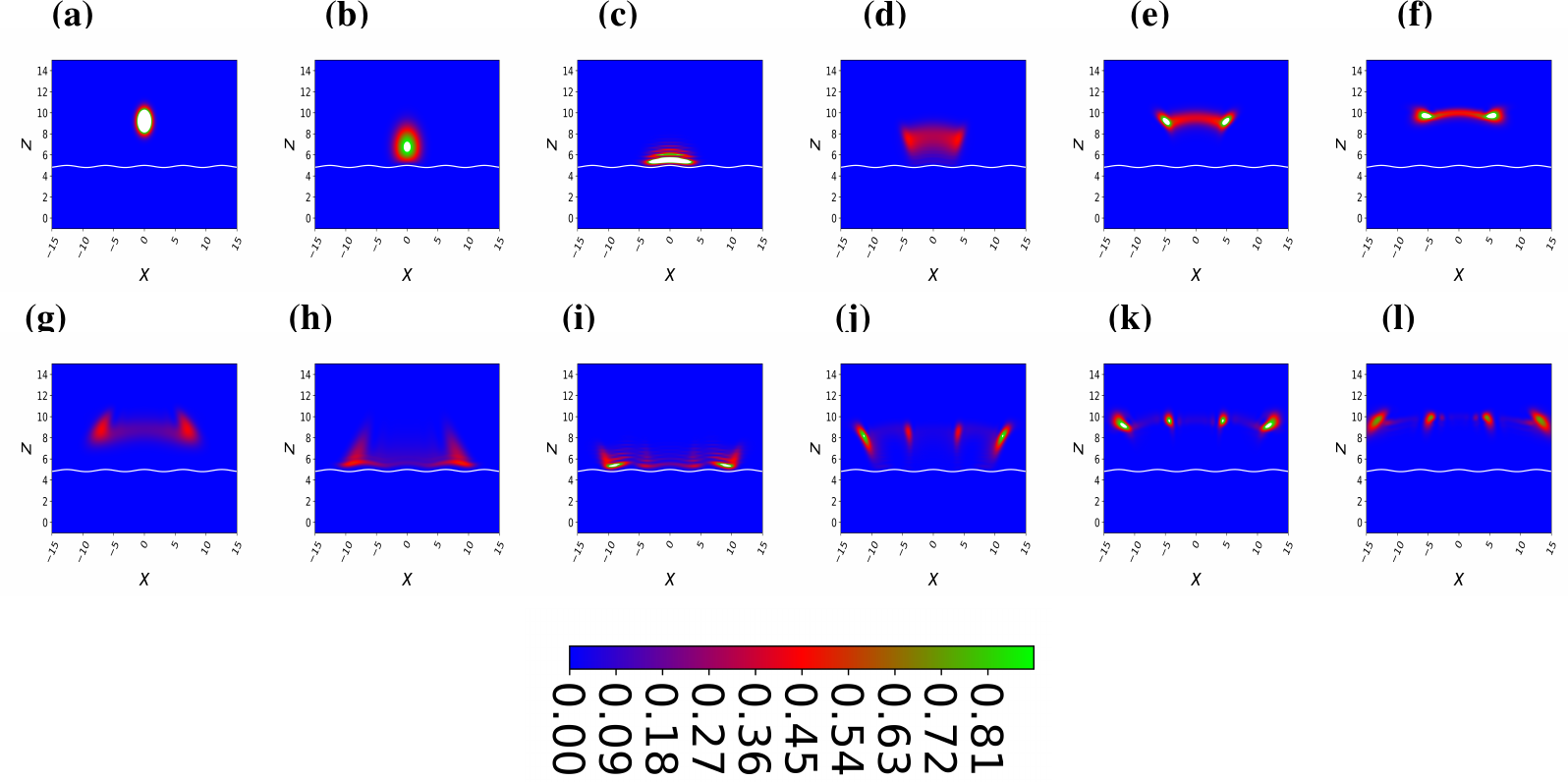}
\end{center}
\caption{(Color online) Snap shots of temporal density of the BEC bouncing under the influence of gravity for different time.  Here, all other dimensionless quantities are  $N=3$, $\eta=0.1$, and $\nu= 1$. Theses images of the retroreflection and diffraction cloud formed by sinusoidal  evanescent wave are taken after every $\triangle t=5$ intervals, we start with $t=5 (a)$ and end with  $t=60 (l)$.  }
\label{fig4}
\end{figure*}

\begin{figure*}
\begin{center}
\includegraphics[width=6.50in,height=3.00in]{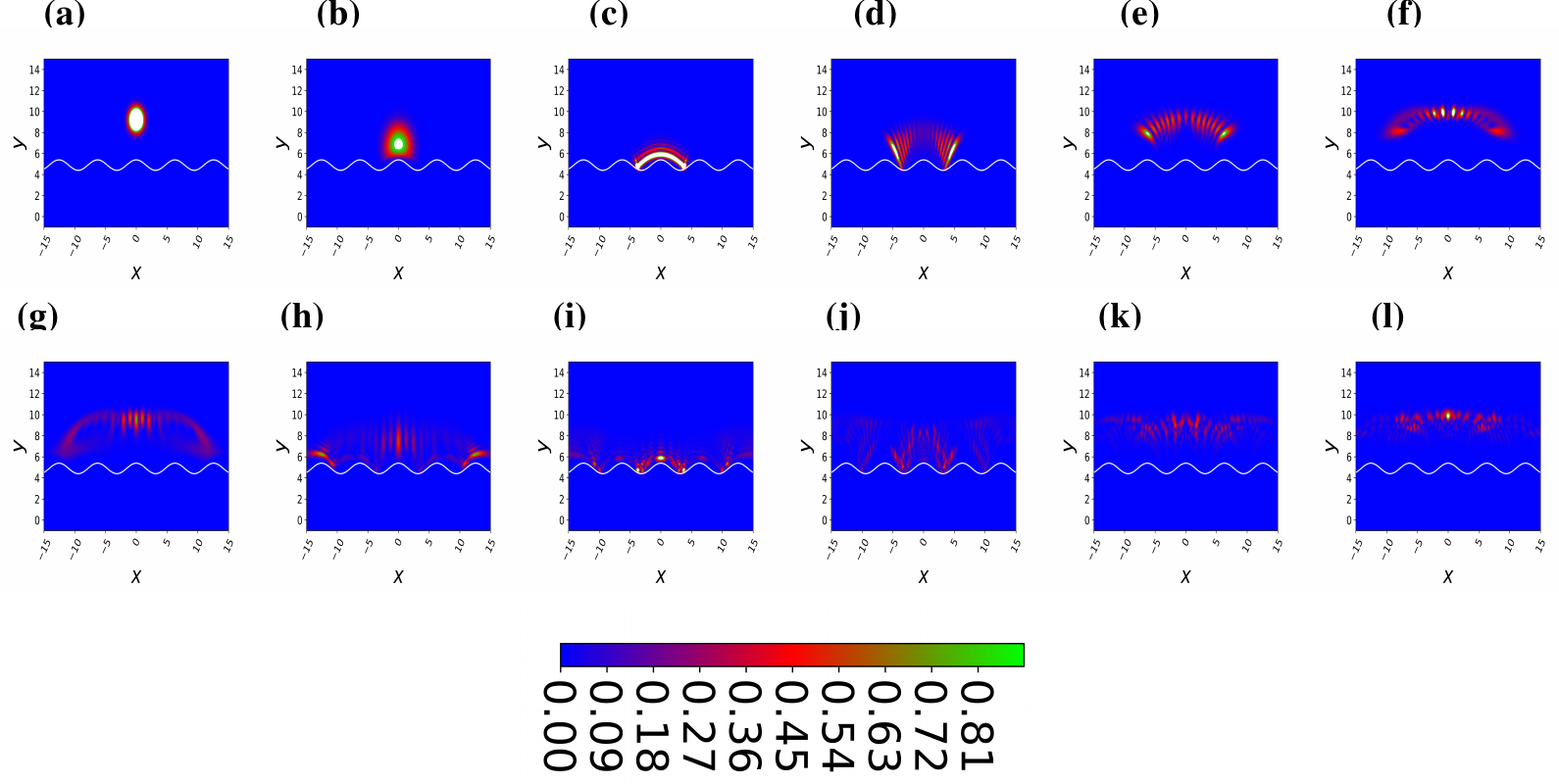}
\end{center}
\caption{(Color online) Snap shots of temporal density of the BEC bouncing under the influence of gravity for different time.  Here, all other dimensionless quantities are  $N=3$, $\eta=0.1$, and $\nu= 1$. These images of the retroreflection and diffraction cloud formed by sinusoidal  evanescent wave are taken after every $\triangle t=5$ intervals, we start with $t=5 (a)$ and end with  $t=60 (l)$.}
\label{fig5}
\end{figure*}

\section{Dynamics of the BEC} \label{Sec3} 
Initially, we trapped a BEC in a two-dimensional trap at the top of the mirror as shown in Fig.  \ref{fig1}(a). Numerically, we achieved this process by letting the initial trapped potential as $V(x,z)= x^2+(z-z_0)^2$, here $z_0=10$. At the time $t=0$, the  initial trapping potential is switched off and condensate is released into the previously described standing evanescent potential  Eq. (\ref{Eq2}) and the BEC  expands ballistically (interacting gas) in potential. Here, numerically we avoid penetration of atoms into the prism by consider a very high potential for the negative $z$-axis i.e., $V(z<0) \rightarrow  \infty $. During the bouncing of the BEC, the atoms experience a weakly confining potential along the $x$-direction and relatively linear along the $z$-direction due to the gravitational field. 

% We notice that for a single particle, the center-of-mass (COM) motion of coincides with the classical trajectory, and follows the dimensionless Ehrenfest theorem, when  $\varepsilon \longrightarrow 0$
% \begin{equation}
%  \frac{d^2 z}{dt^2} =-k + V_0 k e^{-z}
% \end{equation}
% 
% which states that the mean-position of a quantum mechanical particle can be traced by quantum mechanical potentials. However, the Ehrenfest theorem fails when the anharmonicities in the external potential increases. 
%  

We demonstrate the diffraction patterns by using simple geometry. In our model, the potential plane is well approximated by a sinusoidal exponentially decaying field, which can be modeled here as a reflective grating for the BEC atoms. For quantum scattering the periodicity interval $d$ leads to Brag scattering  for constructive interference 

\begin{equation}
 n \lambda_{db}= d \sin(\psi)  , \label{Eq4}
\end{equation}  
with the azimuthal angle  $\psi$ and the de Broglie wavelength 
$ \lambda_{db} =(\hbar/ (2 m E_0))^{1/2} $. Above equation describes the angular positions of diffraction spots $\psi$ of order $n$.  For small angle approximation $\sin(\psi) \approx \psi $, the Bragg condition is given by $\triangle\psi \approx \lambda_{db}/d $, which has been verified experimental for the Fast Atom diffraction \cite{SCHULLER2012161}.  The azimuthal exit angle $\psi$ is associated to the deflection angle $\theta$ in the detection plane  via 
$\tan (\psi) = \tan(\phi_{in}) \sin(\theta)$ as shown in Fig. \ref{fig1}(c), by using the small angle approximation $\sin(\psi)= \sin(\theta) \sin(\phi_{in})  $ the above Eq. (\ref{Eq4}) takes the form

\begin{equation}
 n \lambda_{db\perp}= d \sin(\theta)  , \label{Eq5}
\end{equation}  

where $\lambda_{db\perp}= \lambda_{db}/\sin(\phi_{in})\gg \lambda_{db}$. Here, $\lambda_{db\perp}$ defines as a de Broglie wavelength with respect to the dynamics of projectiles in a plane normal to axial strings as shown in Fig. \ref{fig1}(c). We notice from   Eq. (\ref{Eq5}) that the azimuthal splitting of diffraction spots $\triangle\psi$ for a given de Broglie wavelength $\lambda_{db}$ is independent of the incident angle $\phi_{in}$. The diffraction spots $\triangle\psi$ helps to obtain the periodicity $d$ of the surface potential.   
The external periodic potential frequency leads to a blazing effect similar to an optical grating, which enhances the intensity of finite diffraction order.

\begin{figure*}
\begin{center}
\includegraphics[width=6.00in,height=2.50in]{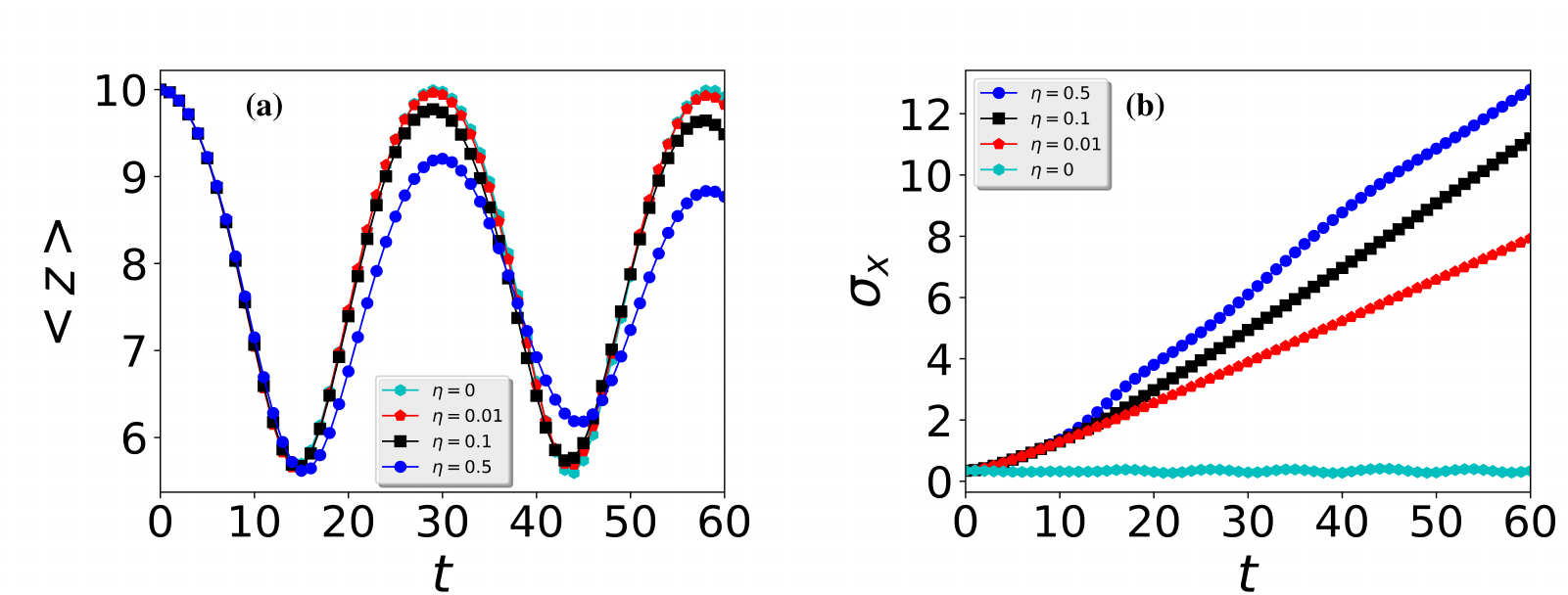}
\end{center}
\caption{(Color online) (a) Dimensionless mean position of the BEC along the z-axis and (b) dimensionless standard deviation $\sigma_x$ of the BEC cloud along the x-axis verses dimensionless time after releasing from the dimensionless height $z=10$ for different periodic potential strength $\eta$. }
\label{fig6}
\end{figure*}

As we already discussed in Sec. \ref{Sec1} that the quantum treatment for the retroreflection and diffraction of a Bose-Einstein condensate by standing evanescent wave does not permit analytical solutions, due to the anharmonicity and nonlinear interaction of BEC atoms. Therefore, we tackle this problem by using a numerical technique called time-splitting spectral method  
\cite{BAO2003318,Vudragovic12,Kumar15,Loncar15,Sataric16},  which perfectly depicts the description of the reflection and diffraction phenomenon.  Therefore, although in this scientific report, we give up analytical approach, but we obtain a detailed insight into the dynamic process of the BEC. The evolution of the BEC probability density as it approaches the evanescent wave and then gets reflected as shown in Fig. \ref{fig2}(a-l). In this special case, the amplitude of a standing wave is zero, therefore, the BEC start bouncing at the exponentially decaying potential. In Fig. \ref{fig2}, at the BEC density graph, additionally we plot a white line that mimics the periodic reflecting potential. The mean position of the BEC exhibits a periodic bouncing as shown in Fig. \ref{fig2}, which is natural as this is the clean case and ideally there are not any impurities present at the prism surface. As we can see that after letting $\eta=0$, our proposed model becomes a one-dimensional case as can be seen from Eq. \ref{Eq2}, therefore to avoid ballistic expansion of the BEC cloud along the x-axis, we explicitly let a harmonic potential $x^2/2$ along the x-axis. If we do not consider the harmonic potential along the x-axis, we notice that we lose all the BEC atoms in the x-axis, therefore to see the retroreflection of the BEC in this special case, we need to let harmonic potential along the x-axis.  This way, we record bouncing of the BEC on a hard surface.  We notice that the width ($\sigma_x$) of the BEC along the x-axis is constant as predicted in Fig. \ref{fig2} and Fig. \ref{fig6}(b), this is quite obvious as we know that along the x-axis the potential is harmonic. We would like to mention that the width of the BEC along the x-axis is model by standard deviation. 

As we make a small increase in the dimensionless amplitude $\eta = 0.01$ of the standing wave as shown in  Fig. \ref{fig3}(a-l),  the BEC temporal density starts to spread during the flight back as can be seen in Fig. \ref{fig3}(d-f). This shows that the prism surface is not smooth anymore, therefore the BEC wavepacket spreads, by studying the spread of the BEC we can predict the shape of the surface structure similar to Surface Physics, where many  experimentalists use the Fast Atom Diffraction technique to study surface patterns \cite{SCHULLER2012161,PhysRevA.94.022711}. After the second reflection from the prism, atoms indicate interference maxima and minima of the temporal density of the BEC as shown in Fig. \ref{fig3}(l) at dimensionless time $t=60$.  This interesting phenomenon can be explained, the interference patterns are produced by overlapping  the part of the BEC wavefunction that is still approaching to the maximum height with the part that is being diffracted previously. These maxima and minima of density wave-packets are quite visible when we further increase the dimensionless amplitude of the standing wave potential, let say $\eta=0.1$ as shown in Fig. \ref{fig4}.  
In the panel Fig. \ref{fig4}(c), the BEC wavefunction is interacting with the standing wave antinode at $x=0$. In this case, there is a circular wave-front formed around the standing wave antinodes, so some of the portion of the wavefunction is diffracted back earlier as compared to the other portion which still touching the minima of the standing wave. The superposition of the circular wave-fronts with each other generates maxima and minima of the temporal density, the temporal density gets weaker and blurred from the middle as we move far from the standing wave as predicted in Fig. \ref{fig4}(f) at dimensionless time $t=30$.  As the BEC wave function further evolves in time, we can see a superposition of four contributions of the BEC wavepacket, as can be seen in the panel (k) of Fig. \ref{fig4} and a rainbow kind of density structure appears. The rainbow pattern has all the information of the surface structure, thus we see four maxima of the density patterns, therefore we can claim that there are four periodic structures are present at the prism surface.  

Interestingly, the BEC wavepacket starts exhibiting the typical Fresnel or near-field features of such functions, these features are very similar to the light when it passes from a diffraction grating. As we increase further the standing wave amplitude to $\eta=0.9$ as shown in  Fig. \ref{fig5}(a-l), the circular wave-fronts of BEC wavepacket reaped around the standing wave antinodes as shown in Fig. \ref{fig5}(c).  We would like to mention here, our mimicking periodic one-dimensional white potential is not describing the exact Physics of the ESWP as gravity starts playing a very important role, therefore, it is getting hard to mimic a two-dimensional potential (\ref{fig1}(b)) as a one-dimensional case. It can be seen from the  density rainbow structure as presented in Fig. \ref{fig5}(e)  that end-part of the BEC wavepacket take more time to reflect back as compared to the middle part. The wavefront of the BEC could not find the perfect plan surface instead the surface is quite rough. So by comparing the clean surface case with the dirty periodic surface case, someone can provide a very good estimate of the periodic surface structure. We note that the mean position of the BEC wavepacket along the z-axis is periodic for all the different dimensionless periodic amplitudes  as shown in Fig. \ref{fig6}(a), however the mean-position of the BEC  decrease with every bounce which means the loss of energy as depicted  in Fig.  \ref{fig6}(a). So a natural question arises where does this energy go, in this closed system, we notice that the total energy of the system is conserved however, this decrease in height is happened due to the spread of the wavepacket in the x-axis. We observe that the standard-deviation (width) of the BEC wavepacket along the x-axis increases linearly with time and the growth of the linear curve increase with the increase of the dimensional periodic strength $\eta$ as shown in Fig. \ref{fig6}(b). For a special case, $\eta=0$ the width of the BEC does not increase as the BEC is bounded in the x-axis due to the harmonic confinement, therefore we do not observe any decrease in mean-position of the BEC wavepacket along the z-axis. For $\eta=0.9$, the BEC wavepacket losses its coherence due to the loss of a definite phase as it bounces on a quite high periodic potential surface and on every bounce a relative phase is added. Therefore,  as the periodic potential strength increases the Quantum decoherence increases and we see quite dull interference patterns as shown in Fig. \ref{fig5}(l), thus for high values of $\eta$, we can not track any information from these interference patterns.

\section{Conclusion}\label{Sec4}
In conclusion, we have demonstrated a simple retroreflection and diffraction technique for ultra-cold atoms, where the diffraction patterns can be controlled with the experimental tuneable parameters. 
We note that the spreading of BEC in the vertical direction is quite arduous. Atoms that originated at the top have a longer bouncing time-period than those started at the center. This effect is quite small during the first few bounces, however, it increases as the number of bounces started to increase. The bright and dark interference fringes can be seen, however, the coherence of the BEC started to lose after few bounces as in very bounce BEC atoms get phase due to path traveling along the z-axis. 
The prospect of exploitation of the effect of rainbow scattering of the BEC wavepacket to examine the structural features of the surface is examined. Fast Atom Diffraction from surfaces is one of the standard methods for solid surface structures near to the room temperature. However, our technique can be used to find out the surface structures at   ultra-cold temperature.  Our technique will open new prospects in the field of BEC diffraction and  reflection from the prism surface. 
Nowadays magnetic-field sensors \cite{BENDING-1999} are efficient of making measurements with both good field sensitivity and  high spatial resolution. For instance, with low sensitivity magnetic force microscopy \cite{Freeman-2001} permits  investigation of magnetic structures with a spatial resolution in the nanometre range, although atomic magnetometers and SQUIDs facilitate to investigate extremely sensitive magnetic-field measurements, but at low resolution. In this respect,  S. Wildermuth et.al., investigated one-dimensional Bose-Einstein condensates in a microscopic field-imaging technique that combines high spatial resolution (within 3 micrometres) with high field sensitivity (300 picotesla) \cite{Wildermuth-2005}. Our proposed model findings can help experimentalists to build future high spatial resolution and high sensitivity magnetic-field sensors in two-dimensional systems.

\section{Method and numerical technique}\label{Sec5}
We derive the Gross-Pitaevskii equation by using the Hamilton principle of least action with the action functional 
\begin{equation}
 S= \int \int \pounds_{3D} dr dt , \label{Eq6}
\end{equation}
where the three-dimensional Lagrangian density is 
\begin{equation}
 \pounds_{3D} = \frac{i \hbar}{ 2} \left( \psi^*_t \psi - \psi_t \psi^* \right)+ \frac{\hbar^2 }{2 m} | \nabla \psi |^2 + V |\psi|^2 + \frac{G_{3D}}{2}  |\psi|^4 . \label{Eq7}
\end{equation}
The dynamics of two-dimensional condensate can be observed when the longitudinal trap frequency $\omega_y$ is much higher than the radial trap frequency $\omega_r$. If this condition holds then the condensate wave function can be factorized 
\begin{equation}
 \psi(r,t)= \phi(y)\psi(x,z,t), \label{Eq8}
\end{equation}
where $\phi(y)= \frac{1}{\pi^{1/4} a_y^{1/2}} e^{\left[-y^2/(2 a_y^2) \right] }$. By substituting Eq. (\ref{Eq8}) into Eq. (\ref{Eq6}) and integrate the over $y$-axis, the effective two-dimensional Lagrangian density 
\begin{equation}
 \pounds_{2D} = \frac{i \hbar}{ 2} \left[ \psi^*_t(x,z,t) \psi(x,z,t) - \psi_t(x,z,t) \psi^*(x,z,t) \right] + \frac{\hbar^2 }{2 m} |\nabla \psi(x,z,t) |^2 + V |\psi(x,z,t)|^2 + \frac{G_{3D}}{2 a_y \sqrt{2\pi}}  |\psi(x,z,t)|^4 . \label{Eq9}
\end{equation}
To determine the two-dimensional time-dependent Gross-Pitaevskii equation, we use the Euler-Lagrangian equation 
\begin{equation}
 \frac{\partial \pounds_{2D}}{\partial \psi^{*}(x,z,t) } -\frac{\partial }{\partial r} \frac{\partial \pounds_{2D}}{  \frac{\partial \psi^* (x,z,t)}{\partial r } } -\frac{\partial }{\partial t} \frac{\partial \pounds_{2D}}{  \frac{\partial \psi^*(x,z,t) }{\partial t } } =0, 
\end{equation}
where, $r=(x,z)$. By using the two-dimensional Lagrangian density Eq. (\ref{Eq9}) and after some algebra, the two-dimensional time-dependent Gross-Pitaevskii equation reads (\ref{Eq1}) \cite{A-2004}.

In order to determine the dynamics of the condensate wave function,
we solve the  two-dimensional time-dependent Gross-Pitaevskii equation (\ref{Eq1}) in imaginary time numerically by using the split-operator method
\cite{BAO2003318,Vudragovic12,Kumar15,Loncar15,Sataric16}. It is worthwhile to mention here that we use the imaginary time simulation technique that helps us to find a suitable ground state wave function, for any random initial guess, for a given interaction strength. The ground state wave function then served as an initial condition for the rest of the dynamic cases. In our case, we use initially Gaussian as a guess which by simulating in imaginary time results in a ground state for a specific interaction strength. For numerical simulations, we perform  discretization of the dimensionless GP Eq. (\ref{Eq3}),
with space step $\vartriangle x= \vartriangle z= 0.083$ and the time step $dt=0.001$.

\section{Acknowledgment}
We gratefully acknowledge support from China Postdoctoral Science Foundation (2018M640440), the Fundamental Research Funds for the Central Universities (3206009604), the National Key R$\&$D program of China (2018YFE0125500,2016YFB0401600, and 2017YFC0111500), National Natural Science Foundation of China (61971133,61674029, 61775034, and 51879042), NSFC Research Fund for International Young Scientists (Grant No. 61750110537).

\bibliographystyle{apsrev4-1}
%\bibliography{differaction.bib}


\end{document}